\documentclass[12pt, referee]{aastex}
\usepackage[dvips]{color}
\newcommand{\re}{\textcolor[rgb]{0,0,0}}

\shorttitle{Terahertz Water Masers}
\shortauthors{Neufeld et al.}

\begin{document}

\title{SOFIA/GREAT$^*$ Discovery of Terahertz Water Masers}
\author{David A.~Neufeld\altaffilmark{1}, Gary J. Melnick\altaffilmark{2}, 
Michael J.\ Kaufman\altaffilmark{3}, Helmut Wiesemeyer\altaffilmark{4}, 
Rolf~G\"usten\altaffilmark{4},
Alex~Kraus\altaffilmark{4}, Karl M.\ Menten\altaffilmark{4}, Oliver Ricken\altaffilmark{4}
and Alexandre Faure\altaffilmark{5}}

\altaffiltext{*}{\re{GREAT is a development by the MPI f\"ur Radioastronomie and the KOSMA/Universit\"at zu K\"oln, in cooperation with the MPI f\"ur Sonnensystemforschung and the DLR Institut f\"ur Planetenforschung.}}
\altaffiltext{1}{Department\ of Physics \& Astronomy, Johns Hopkins University\,
3400~North~Charles~Street, Baltimore, MD 21218, USA}
\altaffiltext{2}{Harvard-Smithsonian Center for Astrophysics, 60 Garden Street, Cambridge, MA 02138, USA}
\altaffiltext{2}{Department of Physics and Astronomy, San Jose State University, One Washington Square, San Jose, CA 95192-0106, USA}
\altaffiltext{4}{Max-Planck-Institut f\"ur Radioastronomie, Auf dem H\"ugel 69, 53121 Bonn, Germany}
\altaffiltext{5}{Univ. Grenoble Alpes, IPAG, and CNRS, F-38000 Grenoble, France}

\begin{abstract}

We report the discovery of water maser emission at frequencies above 1 THz.  Using the GREAT instrument on SOFIA, we have detected emission in the 1.296411 THz $8_{27}-7_{34}$ transition of water toward three oxygen-rich evolved stars: W~Hya, U~Her, and VY~CMa.  An upper limit on the 1.296~THz line flux was obtained toward R~Aql. Near-simultaneous observations of the 22.23508~GHz $6_{16}-5_{23}$ water maser transition were carried out towards all four sources
\re{using the Effelsberg 100m telescope}.  \re{The measured line fluxes imply 22 GHz / 1.296 THz photon luminosity ratios of 0.012, 0.12, and 0.83 respectively for W~Hya, U~Her, and VY~CMa, values that confirm the 22 GHz maser transition to be unsaturated in W~Hya and U~Her.}    We also detected the 1.884888 THz $8_{45}-7_{52}$ transition toward W~Hya and VY~CMa, and the 1.278266 THz $7_{43}-6_{52}$ transition toward VY~CMa.  \re{Like the 22 GHz maser transition, all three of the THz emission lines detected here originate from the ortho-H$_2$O spin isomer.}
Based upon a model for the circumstellar envelope of W~Hya, we estimate that stimulated emission is responsible for $\sim 85 \%$ of the observed 1.296~THz line emission, and thus that this transition may be properly described as a terahertz-frequency maser.  In the case of the 1.885 THz transition, by contrast, our W~Hya model indicates that the observed emission is dominated by spontaneous radiative decay, even though a population inversion exists.

\end{abstract}


\section{Introduction}

Maser action is widely-observed in Nature.   It amplifies the emission from specific transitions of abundant astrophysical molecules in a variety of environments, including the interstellar medium, circumstellar envelopes, and the accretion disks within active galactic nuclei.
Among the half-dozen astrophysical molecules that have been observed to exhibit the maser phenomemon, water vapor possesses the brightest known maser transitions, with brightness temperatures for its 22~GHz $6_{16}-5_{23}$ transition often exceeding $10^{10}$~K and in exceptional cases reaching 10$^{14}$~K.  Thanks to these extraordinarily high brightness temperatures, this transition can be observed by means of Very Long Baseline Interferometry (VLBI), providing \re{submilliarcsecond} angular resolution that enables a variety of astronomical studies that have greatly expanded our understanding of the kinematics of protostellar outflows; helped elucidate the size, shape and kinematics of the Milky Way (e.g. Reid et al. 2009);  and have provided among the best evidence yet obtained for the existence of supermassive black holes (e.g. Miyoshi et al. 1995). 

\begin{figure}
\includegraphics[width=10 cm, angle = 90]{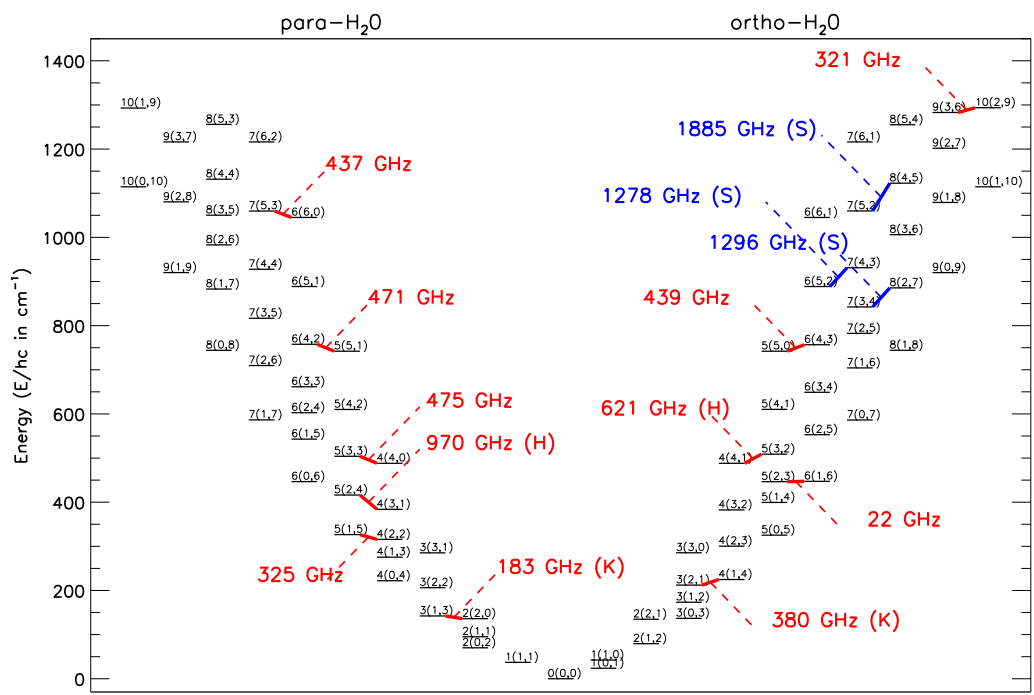}
\caption{Energy level diagram for the ground vibrational state of water. Red lines: maser
transitions observed previously.  Blue lines: THz transitions studied here.  The labels ``K'', ``H'' and ``S'' indicate discoveries obtained using the Kuiper Airborne Observatory, {\it Herschel}, and SOFIA (present study), respectively.  \re{Each state is labeled with the quantum numbers $J(K_{\rm A},K_{\rm C}).$}}
\end{figure}

\begin{deluxetable}{ccccc}

\tablewidth{0pt}
\tabletypesize{\scriptsize}
\tablecaption{Observed maser transitions within the ground vibrational state of water} 
\tablehead{Transition & Frequency & $E_U/k$ $^a$ & Discovery & Environment(s)$^b$ \\
					  & (GHz) & (K) & reference }
\startdata

$6_{16}-5_{23}$ &  22.235 &  644 & Cheung et al.\ (1969) & IS, CSE, AGN \\
$3_{13}-2_{20}$ & 183.310 &  205 & Waters et al.\ (1980) & IS, {CSE, AGN} \\
$4_{14}-3_{21}$ & 380.197 &  323 & Phillips et al.\ (1980) & IS  \\
$10_{29}-9_{36}$& 321.226 &  1861 & Menten et al.\ (1990a) & IS, CSE, AGN\\
$5_{15}-4_{22}$ & 325.153 &  575 & Menten et al.\ (1990b) & IS, CSE\\
$7_{53}-6_{60}$ & 437.347 &  1525 & Melnick et al.\ (1993) & CSE \\
$6_{43}-5_{50}$ & 439.151 &  1089 & Melnick et al.\ (1993) & IS, CSE\\
$6_{42}-5_{51}$ & 470.889 &  1090 & Melnick et al.\ (1993) & IS, CSE\\
$5_{33}-4_{40}$ & 474.689 &  725 & Menten et al.\ (2008) & CSE, IS \\
$5_{32}-4_{41}$ & 620.701 &  732 & Harwit et al.\ (2010) & IS, CSE\\ 
$5_{24}-4_{31}$ & 970.315 &  599 & Justtanont et al.\ (2012) & CSE\\ 
$7_{43}-6_{52}$ & 1278.266 & 1340 & Present work & CSE\\
$8_{27}-7_{34}$ & 1296.411 & 1274 & Present work & CSE \\
$8_{45}-7_{52}$ & 1884.888 & 1615 & Present work & CSE \\
\multicolumn{5}{l}{$^a$ Energy of upper state (temperature units) relative to the ground state of para-H$_2$O} \\
\multicolumn{5}{l}{$^b$ IS = interstellar; CSE = circumstellar envelopes; AGN = active galactic nuclei} \\
\enddata

\end{deluxetable}

In the decades following the first detection of interstellar water vapor through its masing 22 GHz transition (Cheung et al.\ 1969), additional water maser transitions have been discovered at higher frequencies.  In Table 1, we provide a list of all observed maser transitions within the ground vibrational state of water that have been detected at high spectral resolution with the use of heterodyne receivers.  NASA's airborne astronomy program played an early role in these discoveries, with the first detections of the 183 GHz (Waters et al.\ 1980) and 380 GHz (Phillips et al.\ 1980) transitions having been obtained using the Kuiper Airborne Observatory.  In the 1990's and 2000's, thanks to the development of heterodyne receivers on ground-based submillimeter telescopes at dry, high-altitude sites, several additional maser transitions were detected in the 300 - 500~GHz range.  More recently, two additional water maser lines -- at frequencies in the 500 - 1000~GHz range -- were detected using the HIFI instrument on {\it Herschel}.  These transitions are indicated in the energy level diagram shown in Figure 1.

Multitransition maser observations provide key constraints on the pumping mechanisms that lead to a population inversion and on the physical conditions in the masing region (Neufeld \& Melnick 1991; hereafter NM91).  All the masing transitions shown in Figure 1 have $\Delta J = -1$ and \re{$\Delta K_A = +1$}, a behavior predicted by models that invoke a combination of collisional excitation and spontaneous radiative decay as the origin of the population inversion.  Although the quantitative predictions of such models do depend upon the details of the collisional rate coefficients, the question of which transitions become inverted is largely determined by the arrangement of the energy levels (e.g. NM91).  Specific transitions become inverted because the upper state has a longer lifetime than the lower state for radiative decay 
via non-masing far-infrared transitions.  Any population inversion inevitably involves a departure from local thermodynamic equilibrium, and the magnitude of the required departure is an increasing function of the transition frequency, as are the spontaneous radiative decay rates for the masing transition.  These considerations raise the following questions: what is the maximum frequency of water transitions in which a population inversion can occur in astrophysical environments?  And, even if a population inversion is present, at what point does spontaneous radiative decay dominate stimulated emission?  

To address these questions, to test the collisional pumping scheme proposed as the origin of water maser emission, and to probe the physical and chemical conditions within circumstellar envelopes, we used the GREAT instrument on SOFIA to search for Terahertz maser transitions toward four oxygen-rich evolved stars: W~Hya, U~Her, R~Aql, and VY~CMa.  The target sources have all been known to exhibit water maser emission in multiple lower-frequency transitions (\re{e.g.}\ Menten \& Melnick 1991). The primary transition of interest was the $8_{27}-7_{34}$ line at 1.296 THz, a transition predicted to be strongly masing (NM91; Gray et al.\ 2016); in addition, we targeted the $8_{45}-7_{52}$ line at 1.885 THz toward all four sources at the same time as the 1.296 THz line, and the $7_{43}-6_{52}$ line at 1.278 THz toward VY~CMa.  The observing strategy and data reduction methods are described in Section 2 below, and the results presented in Section 3.  In Section 4 we discuss the results in the context of models for water maser emission from the circumstellar envelopes of evolved stars.

\section{Observations and data reduction}

Table 2 lists the evolved stars toward which the water transitions were observed, along with the positions targeted, the estimated distance to each source, the period of the stellar variability, the spectral type, the source systemic velocity relative to the Local Standard of Rest (LSR), and the estimated mass-loss rate.  Also shown are the date of each SOFIA observation, the corresponding stellar phase, the velocity of Earth's atmosphere relative to the LSR, the observatory altitude at the time of the observation, and the total integration time.  

The 1.278 and 1.296 THz water lines were observed in the upper sideband of the L1 receiver of GREAT using the FFT4G backend; the latter provides \re{$\sim 16,000$ spectral channels 
with a spacing of 283~kHz.}
The 1.885 THz water line was observed simultaneously in the lower sideband of the LFA receiver.   
At frequencies of 1.296 and 1.885 THz, the telescope beam has a diameter of $\sim 20^{\prime\prime}$ and $\sim 14^{\prime\prime}$ (HPBW) respectively.  The observations were performed in dual beam switch mode, with a chopper frequency of 1 Hz and the reference positions located 60$^{\prime\prime}$ on either side of the source along an east-west axis.

% LSR versus 
% Earth velocity
% Meaning of integration time

The raw SOFIA data were calibrated to the $T_A^*$ (``forward beam brightness temperature'') scale, using an independent fit to the dry and the wet content of the atmospheric emission.  Here, the assumed forward efficiency was 0.97 and the assumed main beam efficiency for the L1 band was 0.67.  The uncertainty in the flux calibration is estimated to be $\sim 20\%$ (Heyminck et al.\ 2012). 
Additional data reduction was performed using CLASS\footnote{Continuum and Line Analysis Single-dish Software.}.   This entailed the removal of a first-order baseline and the rebinning of the data to a channel width of 0.51~$\rm km \, s^{-1}$ (1296 GHz line), 0.93~$\rm km \, s^{-1}$ (1885 GHz line), or 1.03~$\rm km \, s^{-1}$ (1278 GHz line).

\re{We used the Effelsberg 100-m telescope
to carry out observations of the 22.23508 GHz $6_{16} - 5_{23}$ transition toward all four sources.
In all cases, these observations were performed within 16 days of the SOFIA observations, and for U~Her and VY~CMa they were carried out on the same day.
The 22~GHz observations were performed with the new secondary focus
receiver, with a beam size of 41$^{\prime \prime}$ (HPBW), which was connected to a FPGA-based
FFT spectrometer providing a frequency resolution of 1.5 kHz.
The data were calibrated by correcting for atmospheric opacity and
the dependence of the telescope gain on elevation. For the conversion
of the observed spectra into Jy, suitable calibration sources like 3C~286,
NGC~7027, etc.\  were observed to determine the telescope's sensitivity (which
was about 1 K/Jy for all observations).} 

\begin{figure}
\includegraphics[width=15 cm]{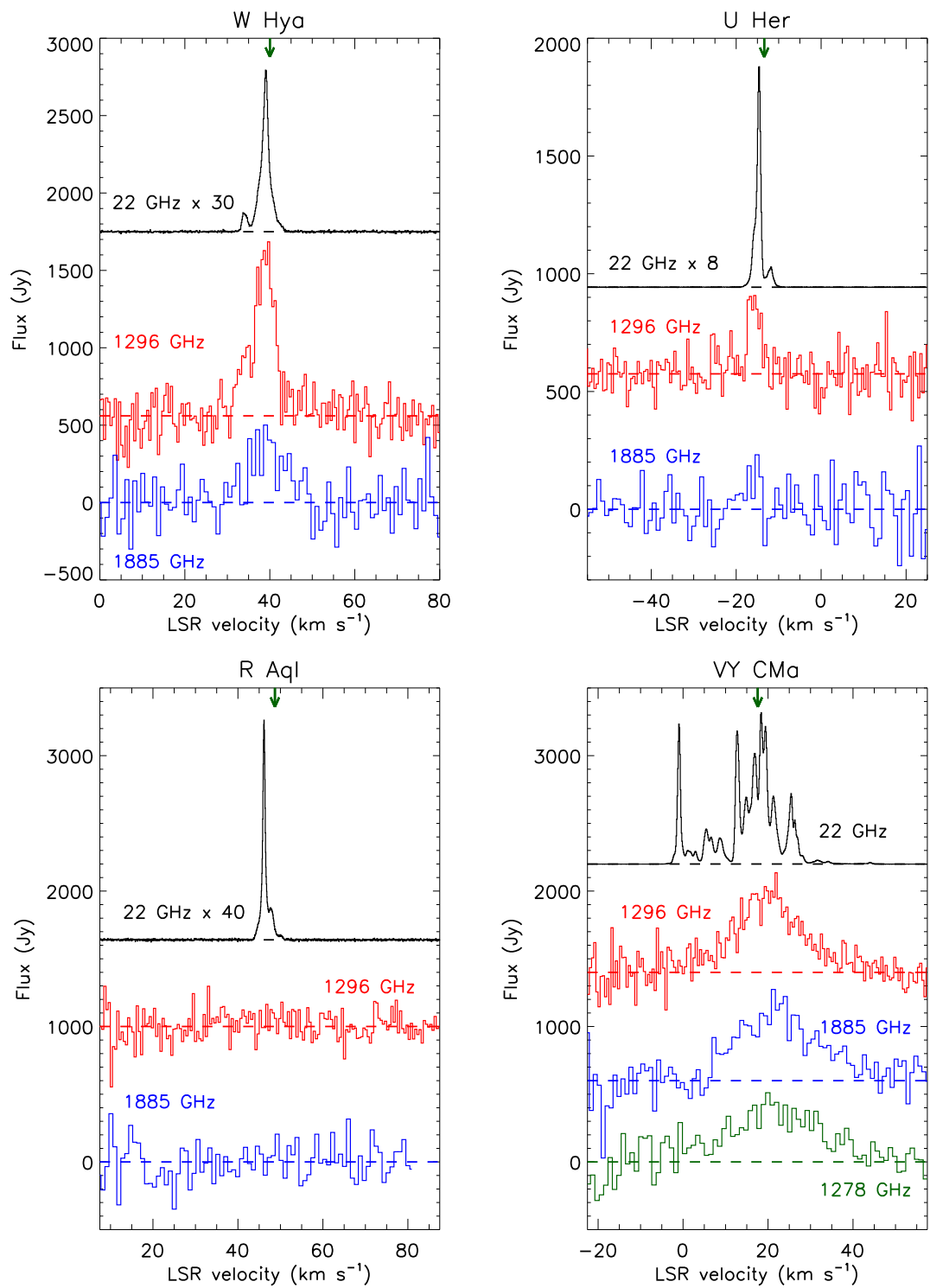}
\caption{Spectra observed toward the four target sources.  Green arrows indicate the systemic velocities of each star.  Vertical offsets have been introduced for clarity, with dashed lines indicating the baseline for each transition.}
\end{figure}

\section{Results}

In Figure 2, we present the reduced spectra obtained from both SOFIA and Effelsberg for all four sources.  The 22 GHz line was observed at very high signal-to-noise ratio in all sources, and the 1.296 THz line was unequivocally detected toward all sources expect R~Aql.  The 1.885~THz line was clearly detected toward W~Hya and VY~CMa.  A search for the 1.278~THz line was only conducted toward one source, VY~CMa, and led to a clear detection.  The 1.278 and 1.296 THz H$_2$O lines had previously been detected toward VY~CMa in spectrally-unresolved observations reported by Matsuura et al.\ (2014) using the SPIRE instrument on {\it Herschel.}  In Table 2, we present the line-integrated fluxes measured for each source, as well as the velocity centroids and velocity dispersions for cases in which a line was detected.

\section{Discussion}

\subsection{Maser line ratios}

For the three stars toward which the 1.296 GHz line was detected, the 22 GHz/1.296 THz photon luminosity ratios listed in Table 2 lie in the range 0.012 to 0.137.  We have compared these values with the predictions of a simple collisional-excitation model in which the steady-state level populations are computed as a function of H$_2$ density {in the range $10^2$ to $10^{12} \, \rm cm^{-3}$}, H$_2$O column density {in the range $10^{10.5}$ to $10^{18.5} \rm \, cm^{-2}$ per km~s$^{-1}$}, and temperature {in the range 10 to 5000~K.}  In this model, the radiative trapping of photons in non-inverted transitions is computed using an escape probability method.  For inverted transitions, the effects of stimulated emission are initially neglected to obtain a solution applicable to the {limit in which the maser gain is small and the population inversion is not reduced by maser emission.}
This yields the (negative) maser optical depths, $\tau({\rm unsat})$, in the unsaturated limit, and also allows us to compute the maximum rate of maser photon emission that can be achieved while maintaining a population inversion.  Full details of the method have \re{been} described by NM91.  The one modification to the treatment discussed in NM91 is the substitution of the best currently-available rate coefficients for the collisional-excitation of H$_2$O by H$_2$; for transitions amongst the lowest 45 rotational states of ortho- and para-H$_2$, we adopt the excitation rate coefficients computed by Daniel et al.\ (\re{2011}), while for transitions involving higher-lying states we use an extrapolation method involving an artificial neural network (Neufeld 2010).  \re{A total of 120 states, all in the ground vibrational state of ortho-H$_2$O, are included.}  Radiative pumping in rotational and rovibrational lines is not included in this model.  Although the infrared radiation field in circumstellar envelopes is typically much higher than in the interstellar medium, calculations by Deguchi \& Rieu (1990) have determined that the excitation of water is still dominated by collisional excitation: in particular, they found that the water line fluxes expected from oxygen-rich evolved stars are typically altered by less than 20$\%$ as a result \re{of} vibrational radiative excitation.\footnote{Our purely collisional pumping model can explain many of the observed relative H$_2$O maser luminosity ratios.  Nevertheless, clearly additional pumping mechanisms must be at work for some of the detected maser lines, most notably for the 437 GHz $7_{53}-6_{60}$ transition. In this line, which has remained undetected in star-forming regions, Melnick et al. (1993) detected strong maser emission toward the Mira variable U~Her with a flux density rivaling that of the 439 GHz line and much higher than that of the 325 and 471 lines (see Table 1). For the Mira star R Leo, Menten et al. (2008) found the 437 GHz to be by far the strongest of the 6 submillimeter H$_2$O maser lines they observed, with a 1200 times higher line luminosity than the 22 GHz line. In their comprehensive study on H$_2$O masers, Gray et al.\ (2016) included {both radiative and collisional} pumping to the $\nu_2=1$ and 2 states of the bending mode. These calculations failed to reproduce the strong maser emission sometimes observed in the 437 GHz line. Apparently, the consideration of other vibrational states or other pumping processes is necessary for reaching a more comprehensive picture of H$_2$O maser pumping in evolved stars.}
{Moreover, in the extensive calculations of water maser emission reported recently by Gray et al.\ (2016), all three of the THz transitions discussed here were found to be in the class of collisionally-pumped masers.}

If both transitions are saturated, the 22 GHz/1.296 THz luminosity ratio is a decreasing function of gas temperature, as expected given the relative energies of the upper states ($E_U/k = 644$~K and 1274~K respectively for the 22 GHz and 1.296 THz transitions).  However, for any combination of gas density, H$_2$O column density, and temperature, the minimum 22 GHz/1.296 THz photon luminosity ratio predicted for saturated maser emission is found to be 0.7, substantially larger than the observed values (0.012 to 0.137) in \re{W Hya and U Her}.  This discrepancy confirms the suggestion, originally made by Menten \& Melnick (1991) on the basis of 22 / 321 GHz \re{maser} line ratios, that the 22 GHz maser transition is \re{often} unsaturated in evolved stars.  This suggestion may also explain the somewhat narrower line widths \re{observed} for the 22 GHz emission from W~Hya and U~Her \re{(see Table 2)}.  
In the case of VY~CMa, although the 22 GHz line shows a velocity dispersion similar to that of the THz lines, the 22 GHz spectrum exhibits more substructure than the THz line spectra.  While a detailed comparison of the measured spectra is limited by the signal-to-noise ratio of the THz observations, it is clear, for example, that the narrow spike at $v_{\rm LSR} \sim -1\,\rm km\, s^{-1}$ in the 22 GHz spectra is absent or relatively much weaker in the THz line spectra.

\begin{figure}
\includegraphics[width=12 cm]{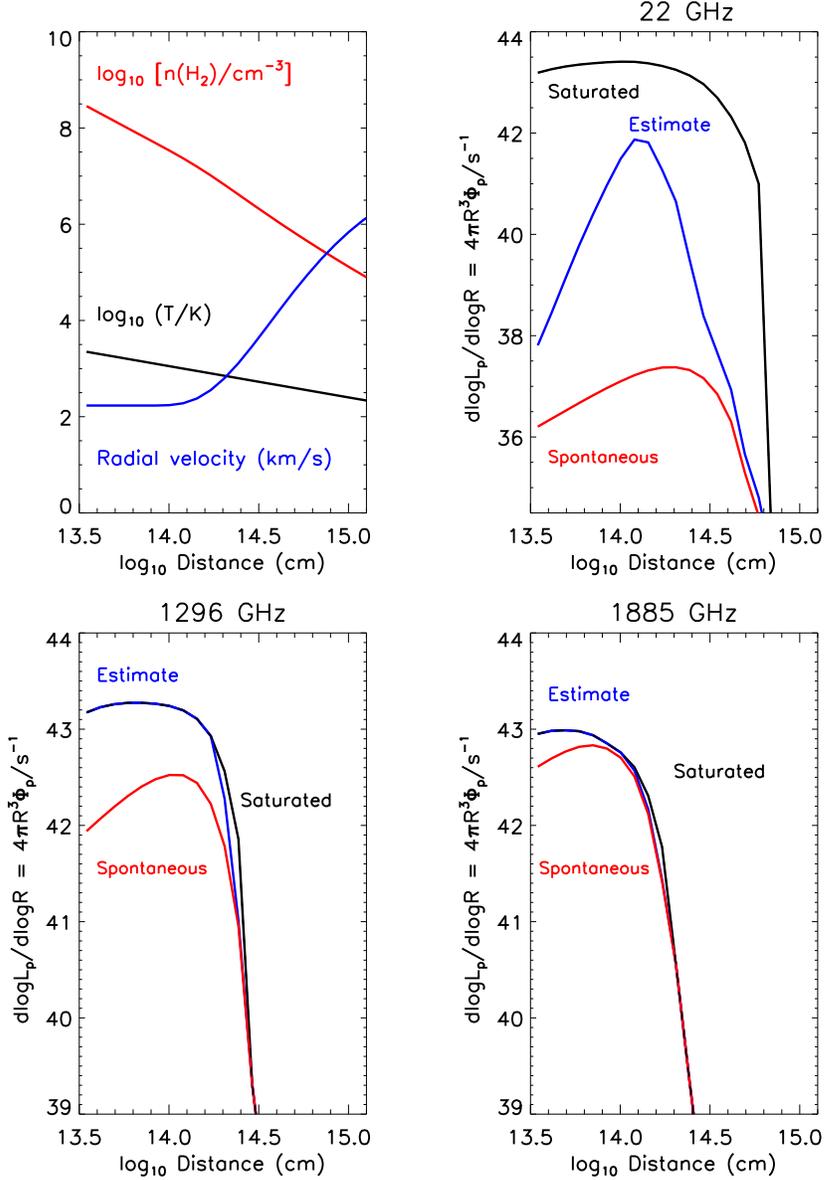}
\caption{Results of a simple model for the water maser emission from W~Hya (see text).  Top left panel: temperature, density, and radial velocity as a function of distance from the center of the star, $R$. Remaining panels (labeled by transition frequency): red curve: rate of spontaneous emission per logarithmic distance element, $4\pi R^3 \Phi_p({\rm spont})$, in units of photons s$^{-1}$. Black curve: total rate of photon emission per logarithmic distance element for saturated maser emission, $4\pi R^3 \Phi_p({\rm sat})$; blue curve: estimate of actual photon emission rate per logarithmic distance element, $4\pi R^3 \Phi_p.$  {The integral of the blue curve $d{\rm log} R$ yields the predicted photon luminosity that is to be compared with the observed value.}}

\end{figure}

\subsection{Simple model for W~Hya}

While the detailed modeling of the observed water maser emissions is beyond the scope of this paper, we present a simplified model for one of the sources, W~Hya.  Here, we adopted the gas temperature and velocity profiles given by Khouri et al.\ (2014), and used the excitation model described above to compute the following quantities for each transition as a function of distance from the star: the 
maser optical depth along a tangential ray in the unsaturated limit, $\tau({\rm unsat})$; the photon emission rate per unit volume in the limit of saturation, $\Phi_p({\rm sat})$; and the rate of spontaneous radiative decay per unit volume $\Phi_p({\rm spont})$.  These quantities were computed with the use of the Large Velocity Gradient (LVG) approximation.   In upper left panel of Figure 3, the gas temperature and radial outflow velocity are shown by black and blue curves as a function of distance, $R$, from the center of the star.  The other three panels, labeled by transition frequency, show $4 \pi R^3 \Phi_p({\rm sat})$ (black) and $4 \pi R^3 \Phi_p({\rm spont})$ (red) for each transition.  Presented in this format, with the volumetric emission rates multiplied by $4 \pi R^3$, these curves show the photon luminosity per logarithmic radial interval.

In regions where a population inversion is present ($\tau({\rm unsat}) < 0$), the total photon emission $\Phi_p$ is bracketed by $\Phi_p({\rm sat})$ and $\Phi_p({\rm spont})$.  For the 22 GHz transition, $\Phi_p({\rm sat})$ typically exceeds $\Phi_p({\rm spont})$ by a factor $\sim 10^6$, and a very large maser gain would be needed to achieve saturation.  For the two THz maser transitions, $\Phi_p({\rm sat})$ exceeds $\Phi_p({\rm spont})$ by at most a factor 10, and only a relatively small gain is needed to achieve saturation.  In our simple model, we estimate the actual photon emission rate, $\Phi_p$, by considering maser amplification along tangential rays, assuming that maser amplifies seed radiation consisting of (1) the cosmic microwave background radiation and (2) radiation emitted by the spontaneous radiative decay.  With these assumptions, we may obtain line flux predictions for all three transitions.  The best agreement with the observed line fluxes, with all three line flux predictions within 15\% of the observed values, was obtained for an assumed mass-loss rate of $\dot{M}=7 \times 10^{-8}\, \rm M_\odot yr^{-1}$ and an ortho-water abundance of $x({\rm o-H_2O})=3.1 \times 10^{-4}$ relative to H$_2$. The resultant gas density profile is shown by the red curve in the upper left panel of Figure 3, and the actual rate of photon emission is shown by the blue curves in the other three panels.  Clearly, for the THz frequency transitions, the predicted line emission is very close to the maximum achievable for saturated masers.  For the 22 GHz transition, however, the emission is unsaturated.  In this case, the predicted line luminosity depends very sensitively upon $\dot{M}$, $x({\rm o-H_2O})$ and the details of the velocity field. 

The best-fit values we obtained for both $\dot{M}$ and $x({\rm H_2O})$ are roughly a factor two below the best-fit estimates obtained by Khouri et al.\ (2014) from a fit to non-masing water transitions.  This discrepancy is perhaps unsurprising given the simple nature of the model.  A more detailed treatment would require the use of a detailed radiative transfer model for the maser radiation and the stellar and dust continuum radiation, with full inclusion of radiative pumping in rotational and rovibrational lines.  Moreover, because the radial velocity gradient is zero within the region interior to the acceleration zone, an accurate calculation would require a treatment beyond the standard LVG approximation.

\subsection{Relative importance of stimulated emission}

In the context of the simple model described above for W~Hya, we may compare the red (spontaneous emission) and blue (estimated total emission) curves in Figure 3 to determine the contribution of maser action to the emergent line fluxes for each of the three observed transitions.  In the case of the 22 GHz transition, spontaneous radiative decay contributes 0.009$\%$ of the total emission, the remaining 99.99$\%$ being contributed by stimulated emission.  For the 1.296 THz transition, the contribution of spontaneous radiative decay is 14$\%$, and that of stimulated emission is $86\%$.  The 1.885~THz transition, by contrast, is dominated by spontaneous radiative decay, which provides $69\%$ of the emergent photons.  These results confirm the declining importance of maser amplification as the transition energy increases.  While the 1.296~THz transition may properly be described as a terahertz maser, the 1.885~THz transition, although inverted, is one \re{for} which maser amplification fails to play a dominant role according to our model for W~Hya.  {This behavior likely reflects the tendency that as the frequency increases, stimulated emission becomes less important relative to spontaneous radiative decay, because the latter typically increases as the cube of the transition frequency.}

\begin{acknowledgements}
Based on observations made with the NASA/DLR Stratospheric Observatory for Infrared Astronomy, \re{and the 100-m radio telescope
of the MPIfR in Effelsberg.} SOFIA Science Mission Operations are conducted jointly by the Universities Space Research Association, Inc., under NASA contract NAS2-97001, and the Deutsches SOFIA Institut under DLR contract 50 OK 0901.   This research was supported by USRA through a grant for SOFIA Program 04-0023.  We gratefully acknowledge the outstanding support provided by the SOFIA Operations Team and the GREAT Instrument Team.

\end{acknowledgements}

{}

\begin{deluxetable}{lcccc}
\tablewidth{0pt}
\tabletypesize{\scriptsize}
\tablecaption{Source parameters, observational parameters, and observed line parameters} 
\tablehead{Source: \phantom{000} & \phantom{000} W~Hya \phantom{000}& \phantom{000}U~Her \phantom{000} & \phantom{000} R~Aql \phantom{000} & \phantom{000} VY~CMa \phantom{000}}
\startdata
  \multicolumn{3}{l}{\underbar{Source parameters}} \\
  \phantom{000000}RA (J2000) & 13h 49m 01.99s & 16h 25m 47.47s & 19h 06m 22.25s & 07h 22m 58.33s\\  
  \phantom{000000}Dec.\ (J2000)) & --28$^o$ $22^{\prime}$ $03.5^{\prime\prime}$ 
   & +18$^o$ $53^{\prime}$ $32.9^{\prime\prime}$ 
   & +08$^o$ $13^{\prime}$ $46.9^{\prime\prime}$ 
  & --25$^o$ $46^{\prime}$ $03.2^{\prime\prime}$ \\
  \phantom{000000}Distance (pc) & $78\pm 6^a$ (K03$^b$) & $266 \pm 30$ (V07) & $214 \pm 39$ (K10) & $1200 \pm 115$ (Z12) \\
  \phantom{000000}Variability period$^c$ (days) 	&  361 & 406 & 270 & 956\\  
  \phantom{000000}Systemic velocity$^d$ (km s$^{-1}$) 		& $40.0 \pm 0.5$	& $-13.4 \pm 0.8$  & $48.7 \pm 0.8$ & $17.6 \pm 1.5$\\
  \phantom{000000}Spectral type$^c$ 				& M7.5e-M9ep & M6.5e-M9.5e & M5e-M9IIIe & M5eIbp(C6,3) \\
  \phantom{000000}Estimated mass-loss rate ($10^{-6}\,M_\odot \rm \, yr^{-1}$) & 0.13 (K14) & 2.6 (G71) & 0.8 (G71) & 150 (R10)  \\
  \\
\multicolumn{3}{l}{\underbar{Observational parameters}} \\  
  \phantom{000000}Date of SOFIA observation & 2016 May 27 & 2016 May 26 & 2016 Nov 10 & 2017 Feb 01\\
  \phantom{000000}Stellar phase$^e$ & 0.71 & 0.99 & 0.64 & ...\\ 
    \phantom{000000}\re{Date of Effelsberg observation} & 2016 May 11 & 2016 May 26 & 2016 Nov 14 & 2017 Feb 01\\
  \phantom{000000}Source velocity relative to SOFIA (km s$^{-1}$) &  $52$ & $-31$ &  $53$ & $41$\\
  \phantom{000000}Observatory altitude (kft) & 43 & 40 -- 41 & \re{39 -- 40.5} & 42\\
  \phantom{000000}SOFIA integration time (min)$^f$ & 11 & 14 & 14 & 23 \\  
  \\
\multicolumn{3}{l}{\underbar{Integrated line fluxes (Jy $\rm km\,s^{-1}$)}} \\
  \phantom{000000}22 GHz      	&  $86.5$ & $167$ & $46$ & $\re{9300}$\\
  \phantom{000000}1.278 THz      &   ...  &  ... & ... & $10500\pm 790$\\  
  \phantom{000000}1.296 THz      &  $6970 \pm 330$ &  $1220 \pm 170$ & $< 1120^g$ & $11220 \pm 520$ \\
  \phantom{000000}1.885 THz      &  $3690 \pm 530$ &  $550 \pm 300 $ & $< 1170^g$& $11080 \pm 660$ \\
  \\
  \multicolumn{3}{l}{\underbar{Photon luminosities ($10^{41}$~s$^{-1}$)}} \\
  \phantom{000000}22 GHz 			&  $3.17$ & $71$ & 12.6 & \re{80700} \\
  \phantom{000000}1278 GHz      &   ...  &  ... & ... & $91100 \pm 6900$\\  
  \phantom{000000}1296 GHz   		&  $255 \pm 12$ &  $519 \pm 73$ & $< 307^g$ & $ 97400\pm 4500$\\
  \phantom{000000}1885 GHz  		&  $135 \pm 19$ &  $235 \pm 128$ & $< 322^g$ & $96100 \pm 5700$\\ 
  \\
  \multicolumn{3}{l}{\underbar{Photon luminosity ratios}} \\ 
  \phantom{000000}22 GHz/ 1296 GHz   & $0.0124 \pm 0.0006$ & $0.137 \pm 0.019$  & $> 0.041$
  & \re{$0.829 \pm 0.0038$} \\
  \\
  \\
\multicolumn{3}{l}{\underbar{Line centroids ($\rm km\,s^{-1}$ relative to the LSR)}} \\
  \phantom{000000}22 GHz 			&	38.7 	& --14.6 &  46.5 & 15.1 \\	
  \phantom{000000}1.278 THz      &   ...  &  ... & ... &  19.7 \\
  \phantom{000000}1.296 THz   		&   38.7 	& --15.4 & ... & 19.6\\
  \phantom{000000}1.885 THz  		& 	39.7    & --16.3 & ... & 22.6\\
  \\
\multicolumn{3}{l}{\underbar{Line {velocity dispersions} (in $\rm km\,s^{-1}$)}} \\
  \phantom{000000}22 GHz 			&	1.85 	& 1.17 &  1.08 & 7.93\\	
  \phantom{000000}1.278 THz      &   ...  &  ... & ... & 10.95\\
  \phantom{000000}1.296 THz   		&   3.26 	& 1.73 & ... & 8.77\\
  \phantom{000000}1.885 THz  		& 	3.66    & 1.68 & ... & 8.12\\
\\
\multicolumn{5}{l}{$^{a}$Where given, errors are 1~$\sigma$ statistical errors} \\
\multicolumn{5}{l}{$^{b}$References: K03 = Knapp et al.\ (2003); V07 = Vlemmings \& van Langevelde (2007); K10 = Kamohara et al.\ (2010);} \\
\multicolumn{5}{l}{\phantom{000}Z12 = Zhang et al.\ (2012); K14 = Khouri et al.\ (2014); G71 = Gehrz \& Woolf (1971); R10 = Royer et al.\ (2010)}  \\
\multicolumn{5}{l}{$^{c}$General Catalogue of Variable Stars, v5.1 (Samus et al.\ 2017)} \\
\multicolumn{5}{l}{$^{d}$Relative to the LSR, as determined from observations of thermal SiO emission (Dickinson et al.\ 1978)} \\
\multicolumn{5}{l}{$^{e}$Determined from recent visual light curves generated by AAVSO (www.aavso.org)} \\
\multicolumn{5}{l}{$^{f}$On source integration time} \\
\multicolumn{5}{l}{$^{g}$3$\sigma$ upper limit obtained for the $43 - 52 \, {\rm km \, s}^{-1} v_{\rm LSR}$ range} \\

\enddata
\end{deluxetable}

\end{document}